\documentclass[acus]{JAC2000}

\usepackage{graphicx}

\begin{document}
\title{\flushright{FRBT004}\\[15pt] \centering MODEL DRIVEN RAMP CONTROL AT RHIC\thanks{Work supported by U.S. DOE under contract No. DE-AC02-98CH10886.}}

\author{J. van Zeijts, Collider-Accelerator Department,\\ Brookhaven National Laboratory, Upton, NY 11973, USA}

\maketitle

\begin{abstract}
At the Relativistic Heavy Ion Collider (RHIC), magnets are ramped from injection energy to storage energy in several minutes
 where it is to remain for several hours. The path up the ramp is marked by 'StepStones' where the  the optics of the machine, which can change dramatically when we perform a beta*-squeeze, is given in units like Quadrupole focusing strength or Corrector-Dipole angle. The machine is tuned at these Stepstones, and at Injection or Storage, by specifying physics properties like Tunes and Chromaticities. An on-line model server handles conversion to magnet strengths, and predicts the optics along the whole ramp.

We will describe the underlying principles, the client-server environment, including on-line model servers, Ramp Manager and Editor, and present operational experience with the system.
\end{abstract}

\section{Introduction}
The RHIC magnets are driven by about 1000 Wave Form Generators (WFG). Most Quadrupole magnets are hooked up through a nested power supply scheme, which minimizes the number of high current cryogenic feed-throughs, but complicates their programming considerably. A more detailed description of the Ramp control is given in~\cite{ramp}, here we concentrate on the physics control and modeling sections.

\section{Magnet Control}
Magnets are programmed in physics units like KL (integrated strength), and angle. The WFG's execute formulas at 720Hz that read the machine magnetic rigidity from the real time data link (RTDL), look up the interpolated requested magnet strength, calculate the required field strength, and use the magnetic transfer table to calculate currents for the associated power-supplies.

\subsection{StepStones}
StepStones are placeholder for a set of magnets and their associated strengths. The strengths is split up in a `Design', and `Trim' part. The machine is set to the design level by modifying the trim settings, client applications usually use the design part of the strength for model calculations, since it more closely resembles the real machine.
StepStones are sparse, in a sense that only some magnets need to be set explicitly, all other are interpolated as a function of the relativistic gamma. The interpolation scheme is critical for proper power supply performance, and involves cubic splines for Quadrupole and Sextupole magnet strengths. Other types of magnets use linear interpolation of strength.

\subsection{Ramps}
\begin{figure*}[t]
\centering
\includegraphics*[width=170mm]{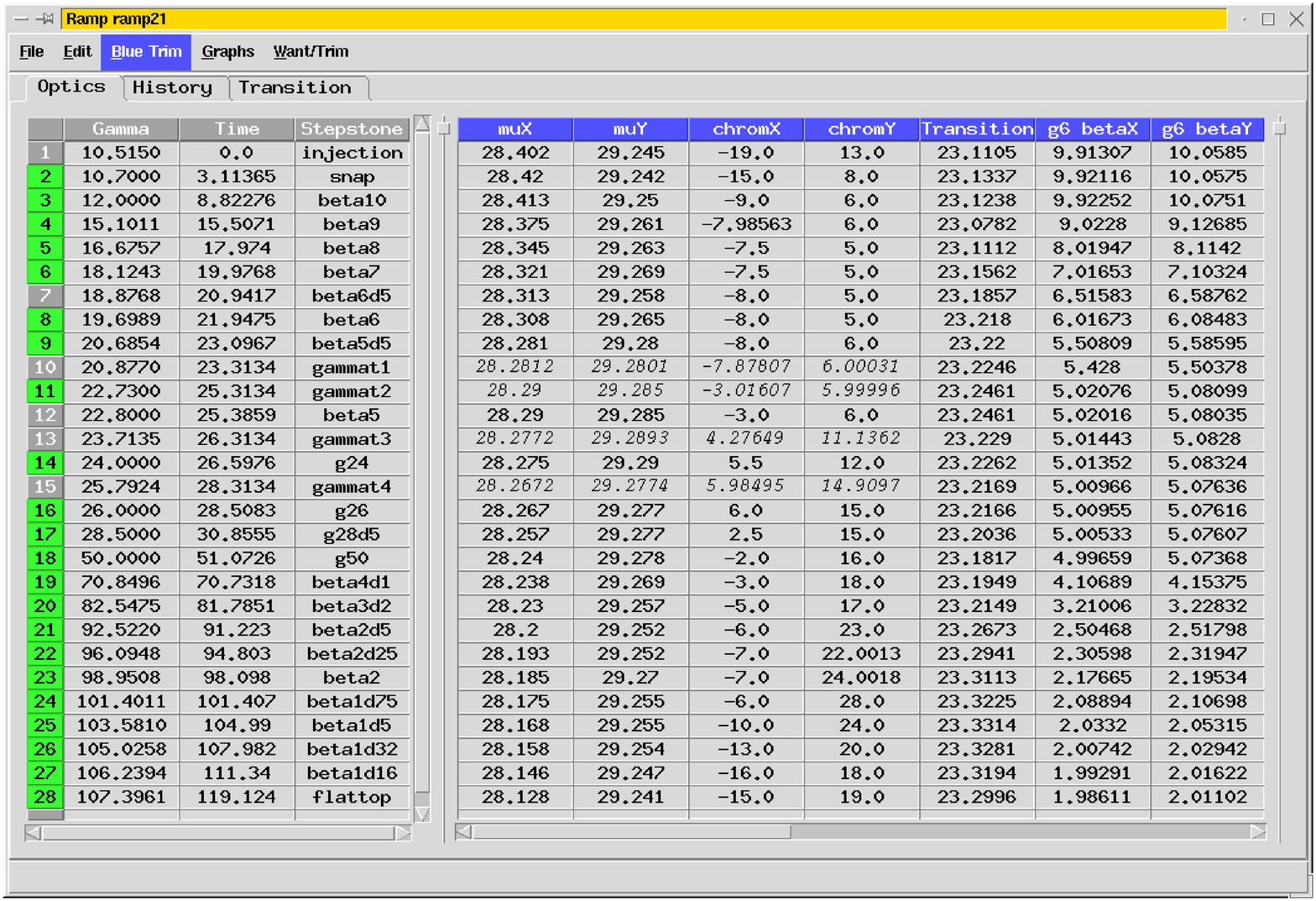}
\caption{High-level display of a ramp in the Ramp-Editor. Tunes and Chromaticities are modified from this page.} \label{rampeditor}
\end{figure*}

\begin{figure*}[t]
\centering
\includegraphics*[width=170mm]{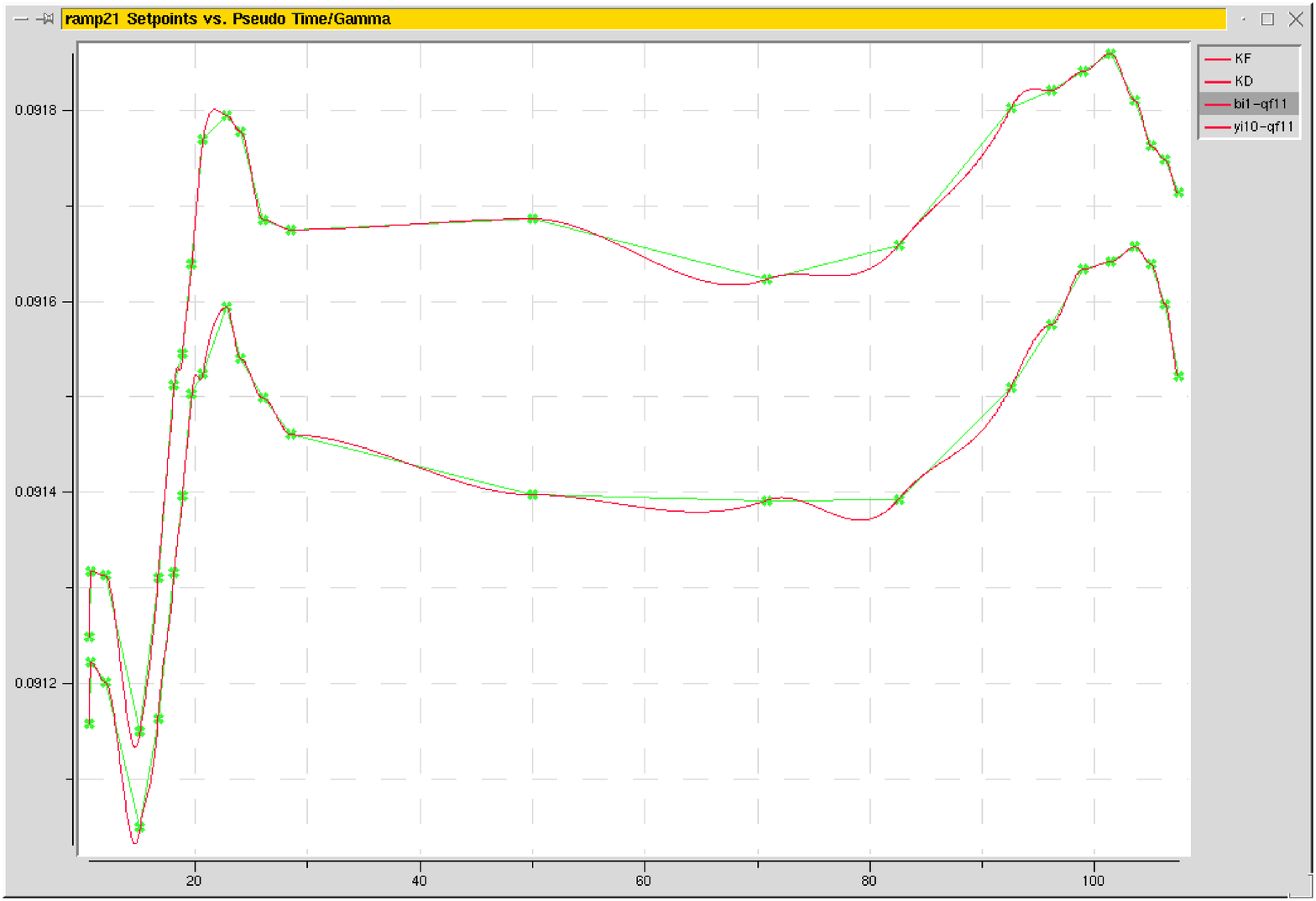}
\caption{Main Quadrupole magnet strength vs. gamma. The Green markers are at the location of the StepStones, the smooth lines are the cubic-spline interpolation.} 
\label{K F/D}
\end{figure*}

Ramps are placeholder for a set of StepStones. The ramps in use at the moment for RHIC accelerate, and Beta*-squeeze at the same time. The model server does optics simulations at many points along the ramp, giving tunes and chromaticity predictions that can be compared with measured numbers. The model can contain multiple named ramps simultaneously, each containing tens of stepstones (see Fig.~\ref{rampeditor} for a typical ramp layout, Fig.~\ref{K F/D} for a graph of the main quadrupole strength).

\section{Model Servers}
Multiple model servers are available, each presenting an identical interface. The differences are in speed and accuracy. The fast model only considers linear un-coupled optics. There are on-line models available which consider full coupling, nonlinearities etc.~\cite{mod1}, but with the associated longer execution time.
For regular machine operation the linear model is preferred, for studies we can switch to a more complete model.

The model server is implemented using the CDEV~\cite{cdev, cdevs, ical} generic server framework, which allows for rich data structures to be passed between client and server. Ramps and StepStones are accessible as CDEV devices, and present properties which can be monitored by client applications. Clients receive updates when magnet strengths are modified. All typical optics properties are exported, the most commonly used ones include:
\begin{Itemize}
\item `LatticeFunctions', clients specify a beam line (Blue or Yellow) and a list of element names. The server by default returns a full set of lattice functions. The context can be modified to only request certain lattice functions.
\item `OpticsFunctions', clients specify a beam line. The server returns a list of tunes, chromaticities, etc.
\item `Orbit', clients specify a beam line, and a list of element names. The server returns the predicted orbit using the dipole corrector set points.
\end{Itemize}

\section{Client Applications}
The on line model server is the hub for lattice and optics information. Magnetic element strength are handled in a separate Ramp-Manager.
Applications routinely retrieve and monitor element strengths and lattice functions at specific StepStones, and (at a higher resolution) along the ramp.
Below is a subset listed of client applications connected to the model.

\subsection{Ramp Editor}
	The main Ramp control GUI allows modification to tunes, chromaticities, and individual element strengths. On each change the model 
	recalculates the predicted optics at each stone, and along the ramp. 

\subsection{Injection Application}
	Injection into both the RHIC rings is facilitated by the `Injection Application'. This application retrieves
	the transverse lattice functions in the transfer line and the first sextant of the rings from the model server.
	Dipole corrections for optimized injection, and closed orbits are calculated and sent to the Ramp Manager. Predicted and measured orbits are displayed.
	
\subsection{Orbit Correction}
	Global Ring Orbit-Correction, Local Correction, 3 and 4 Bump construction etc. are 
	supported in this application. Dipole correctors strengths are calculated and set though this application. Lattice function information, 
	including phase advance between correctors and Beam Position Monitors (BPM) is retrieved from the model. Predicted and measured orbits are displayed.

\subsection{Transverse Profile Manager}
	Lattice functions at the Profile pickups are monitored by the 'Profile-Manager', measured profiles are then converted to 	normalized emittance at injection, up the ramp, and at storage energies.
	
\subsection{Luminosity Monitor}
	Beta functions at the interaction regions are monitored by the 'Luminosity-Monitor', which combines this information with beam intensity and compares measured and predicted luminosity.

\subsection{Coupling-Correction Application}
	In order to correct transverse coupling in the machine the tune set-points are swept over a given range while plotting the measured tunes vs. set-points..
	The correction application utilizes the model to calculate the required set points.

\subsection{Sequencer}
	Progress though the many steps required to run the RHIC through its machine cycle is choreographed by the 'Sequencer' program~\cite{seq, seq1}. This program sets the `liveRamp' and `liveStone' CDEV devices to their appropriate value during the cycle. Client applications usually use these aliases to get updates on the current optics, instead of named stepstones.

\section {Operational Experience}
Having a consistent source of optics information is critical for commissioning a complex machine. The on-line model servers provide such a source. The servers have been in operational use for several years serving client applications routinely used to run the machine. The interface to the servers is through a well defined CDEV interface, which much simplifies the client application programming. The system of servers is flexible, and performs reliably even under simultaneous load of tens of client applications.

\end{document}